\documentstyle[11pt,newpasp,twoside]{article}
\input{epsf}

\def\edcomment#1{\iffalse\marginpar{\raggedright\sl#1\/}\else\relax\fi}
\reversemarginpar

\newcommand  \kms      {\ifmmode {\rm km\,s}^{-1} \else km\,s$^{-1}$\fi}
\newcommand  \cmii     {\hbox{cm$^{-2}$}}
\newcommand  \ergcms   {\ifmmode {\rm erg\,cm}^{-2}\,{\rm s}^{-1}
                        \else erg\,cm$^{-2}$\,s$^{-1}$\fi}
\newcommand  \mr       {MR~2251$-$178}

\begin{document}
\title{The Properties and the Evolution of the Highly Ionized Gas in MR~2251$-$178}
\author{Shai~Kaspi, Hagai~Netzer, Doron~Chelouche}
\affil{School of Physics and Astronomy, Raymond and Beverly Sackler
Faculty of Exact Sciences, Tel-Aviv University, Tel-Aviv 69978, Israel}
\author{Ian~M.~George, T.~J.~Turner}
\affil{Joint Center for Astrophysics, Physics Department, University
of Maryland, Baltimore County, 1000 Hilltop Circle, Baltimore, MD 21250, USA,
and Laboratory for High Energy
Astrophysics, NASA/ Goddard Space Flight Center, Code 662, Greenbelt,
MD 20771, USA.}
\author{Kirpal~Nandra}
\affil{Astrophysics Group, Imperial College London, Blackett 
Laboratory, Prince Consort Road, London SW7 2AZ, UK.}

\begin{abstract}
We present the first {\it XMM-Newton} observations of the radio-quiet
quasar \mr . We model the X-ray spectrum with two power laws, one at
high energies with a slope of $\Gamma=1.6$ and the other to model
the soft excess with a slope of $\Gamma=2.9$, both absorbed by at
least two warm absorbers (WAs). The high-resolution grating spectrum
shows emission lines from N\,{\sc vi}, O\,{\sc vii}, O\,{\sc viii},
Ne\,{\sc ix}, and Ne\,{\sc x}, as well as absorption lines from the
low ionization ions O\,{\sc iii}, O\,{\sc iv}, and O\,{\sc v}. A
study of the spectral variations in \mr\ over a period of 8.5
years yields that all X-ray observations can be fitted with the
above model. Luminosity variations over timescales of years seem
to correlate with the soft excess variations but not with the WA
properties variations. The overall picture is that of a stratified
WA that enters and disappears from the line-of-sight on timescales
of several months.  We also present the first {\it FUSE} spectrum of
\mr . The general characteristics of the UV and X-ray absorbers seem
to be consistent.
\end{abstract}

\section{Introduction}

\mr\ ($z=0.06398\pm0.00006$, $V\approx 14$) is the first quasar
detected by X-ray observations (using $Ariel \ V$ and $SAS$-3 in 1977)
and also the first quasar where a warm absorber (WA) was suggested
to explain the X-ray spectrum (Halpern 1984). It was observed by
practically all X-ray missions since $Einstein$. Previous X-ray
data of \mr\ were modeled using a power-law with photon index
$\Gamma\approx 1.6$--1.7 and a WA with column density in the range
$10^{21.3-22.2}$\,\cmii (e.g., Mineo \& Stewart 1993; Orr et al. 2001;
Morales \& Fabian 2002).
The X-ray flux of the source is variable on timescales of $\sim
10$ days. The observed 2--10 keV flux of \mr\ covers the range of
$(1.7$--$5.1)\times10^{-11}$ \ergcms\ which translates to a 2--10
keV luminosity of $(1.7$--$5.1)\times10^{44}$~\ergcms .

UV spectra of \mr\ were obtained by the {\it Hubble Space
Telescope} at three epochs and show clear Ly$\alpha$ and C\,{\sc iv}
absorption. Ganguly et al. (2001) found the C\,{\sc iv} doublet
absorption to vary with time, suggesting an intrinsic origin for
this absorption.

In this contribution we present new {\it XMM-Newton} and {\it FUSE}
observations of \mr . We also carry out an in-depth analysis of
the 10 available {\it ASCA} observations and the two {\it BeppoSAX}
observations.

\begin{figure}[t]
\centerline{
\hglue0.0cm{\epsfxsize=2.5in\epsfbox{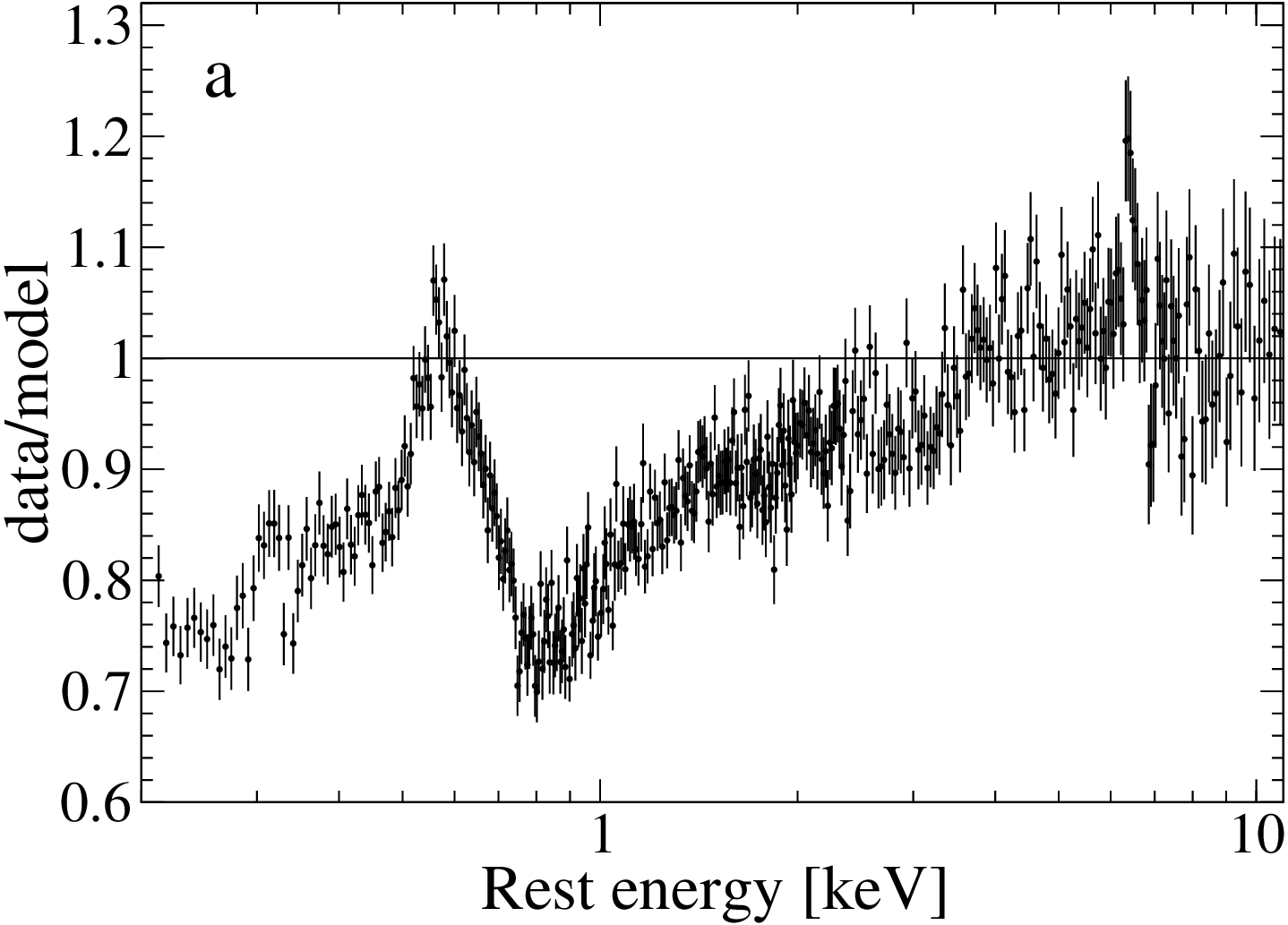}}
\hglue0.1cm{\epsfxsize=2.5in\epsfbox{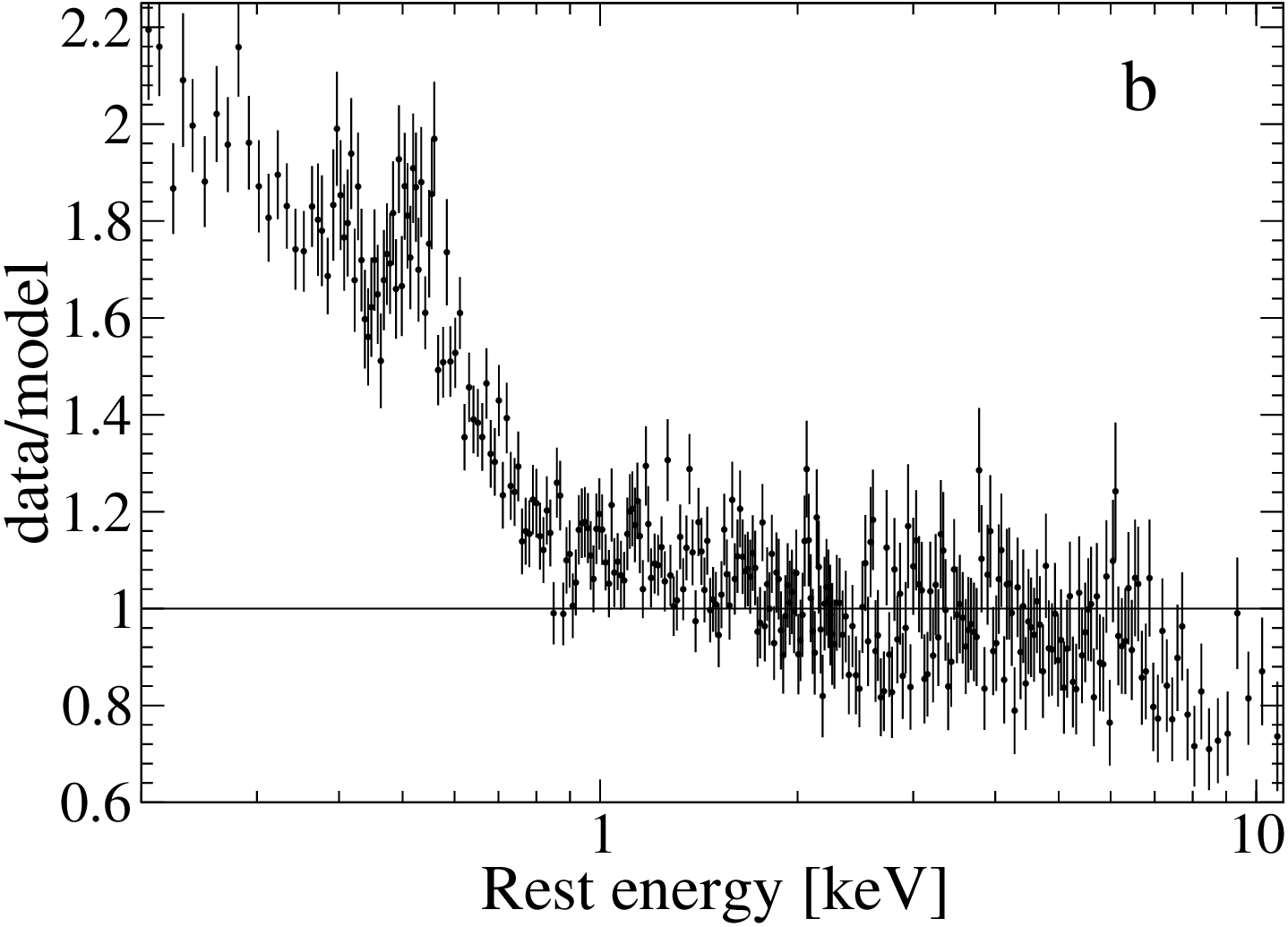}}
}
\caption{(a) Ratio of the 2002 {\it XMM-Newton} EPIC-pn data to a
power law model (including Galactic absorption of $10^{20.45}$\,\cmii)
fitted to the 3--11 keV band. Excess absorptions is evident near the
O\,{\sc vii} and the O\,{\sc viii} absorption edges and at energies
below 0.5 keV. (b) Ratio of the 2000 {\it XMM-Newton} EPIC-pn data
to the scaled 2002 EPIC-pn model.  The additional soft excess in the
2000 spectrum is evident.}
\vskip -0.15in
\end{figure}

\section{The X-ray Spectra and Past variations}

\mr\ was observed with $XMM$-$Newton$ in 2000 and in 2002.  The flux
in the 2002 observation was about 2.5 times lower than the 2000
observation. The EPIC-pn data of the 2002 observation clearly show
a power law with a photon index of $\Gamma\approx 1.6$ at energies
above 3 keV. Extrapolating this power law to lower energies revels the
presence of a WA
around 0.8 keV, an additional absorber below 0.5 keV, and some excess
emission around 0.5 keV (Figure 1a). Our best fitted model for these
data yields a WA with a column density, $N_{\rm H}$, of $10^{21.51\pm0.03}$
\cmii, ionization parameter, $U_{\rm OX}$, of $10^{-1.78\pm0.05}$ and
a line of sight covering factor of 0.8. Assuming gas with the
same properties produces the emission, we find a global covering factor
of 0.3. For the less ionized absorber we find that it can be fitted
by a neutral absorber (in addition to the galactic one) with $N_{\rm H}
\approx 10^{20.3}$ \cmii . This absorber can also be modeled as a
combination of low-ionization absorber with $\log(U_{\rm OX}) \approx -4$
and $N_{\rm H} \approx 10^{20.3}$ \cmii\ and a neutral absorber of $N_{\rm H}
\approx 10^{20.06}$ \cmii . Both cases give equally good fits.
Fitting the above model to the 2000 observation clearly revels the
presence of a soft excess (Figure 1b) which can be fitted with an
additional power law with a photon index of $\Gamma\approx 2.9$
at energies less than $\sim 1$ keV.

The high-resolution grating spectrum of the 2002 observation is
shown in Figure 2. The spectrum shows emission lines from N\,{\sc
vi}, O\,{\sc vii}, O\,{\sc viii}, Ne\,{\sc ix}, and Ne\,{\sc x}, as
well as absorption lines from the low ionization ions O\,{\sc iii},
O\,{\sc iv}, and O\,{\sc v}.  The data were fitted with a detailed
photoionization model which includes the components described above
and is shown in Figure 2.

\begin{figure}[t]
\centerline{\epsfxsize=5.2in\epsfbox{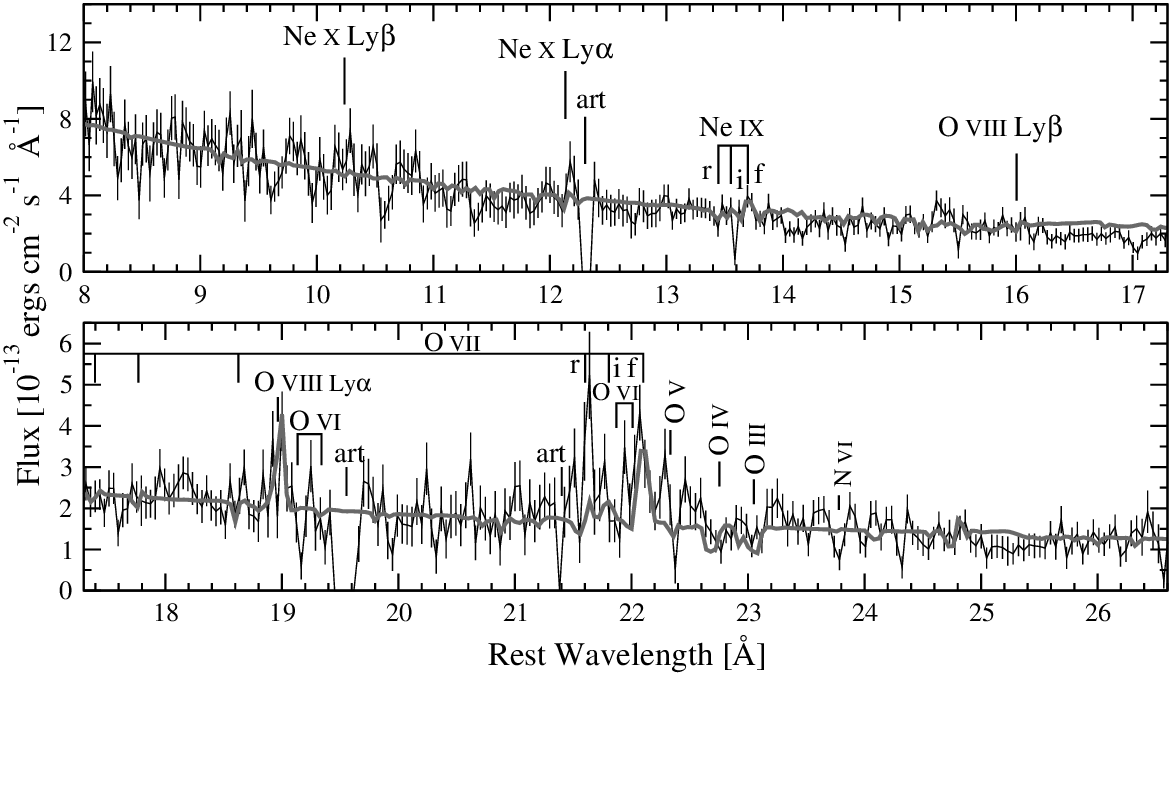}}
\vskip -0.7in
\caption{Combined RGS1 and RGS2 spectrum of \mr\ binned to $\sim 0.04$
\AA . The spectrum has been corrected for Galactic absorption and for
the redshift of the source. The strongest emission lines are due to
the O\,{\sc vii} triplet and O\,{\sc viii} Ly$\alpha$. Other suggested
absorption and emission lines are marked.  Gaps in the spectrum due
to chip gaps are marked as `art'. The two absorbers model described
in the text is shown (gray line).}
\vskip -0.15in
\end{figure}

We used the model with the two power laws and two absorbers described
above to fit historical data of \mr\ obtained over 8.5 years. The WA
properties during the 6 weeks of $ASCA$ observations in 1993 were
consistent with those of an absorber with a column density of
$10^{21.5}$ \cmii . These data are consistent with the scenario in
which the decrease in flux caused a corresponding decrease in the
ionization parameter. However, the \mr\ observations are not frequent
and detailed enough to infer on the location of the gas.

On timescales of years, the WA properties change in time but are
not correlated with luminosity variations. Our only successful model
requires that the absorbing material is changing in time. For example,
the $ASCA$ 1996 observations clearly indicate a larger column density
($>10^{21.8}$ \cmii\ vs. $10^{21.5}$ \cmii ) and a smaller ionization
parameter ($\log U_{\rm OX} \sim -2.3$ vs. $\sim -2.0$) of the WA
compared to 1993. We suggest that a physical motion of gas into and
out of our line-of-sight can cause these changes in the absorber
properties.
In the 2002 $XMM$-$Newton$ observation, the flux is smaller by a
factor of 2 compared with 1993 $ASCA$ observations, yet the column
density and the ionization parameter are similar. Comparing the two
$XMM$-$Newton$ observations yields a similar conclusion. This might
mean that the absorbing material has changed between the
epochs since the SED is very similar but the luminosity decrease
between the epochs was not accompanied by a corresponding decrease
in ionization parameter. An alternative explanation
can be that the absorbing material is far from the central source
and does not respond to the luminosity variations.

\begin{figure}[t]
\hglue-0.5cm{\epsfxsize=5.4in\epsfbox{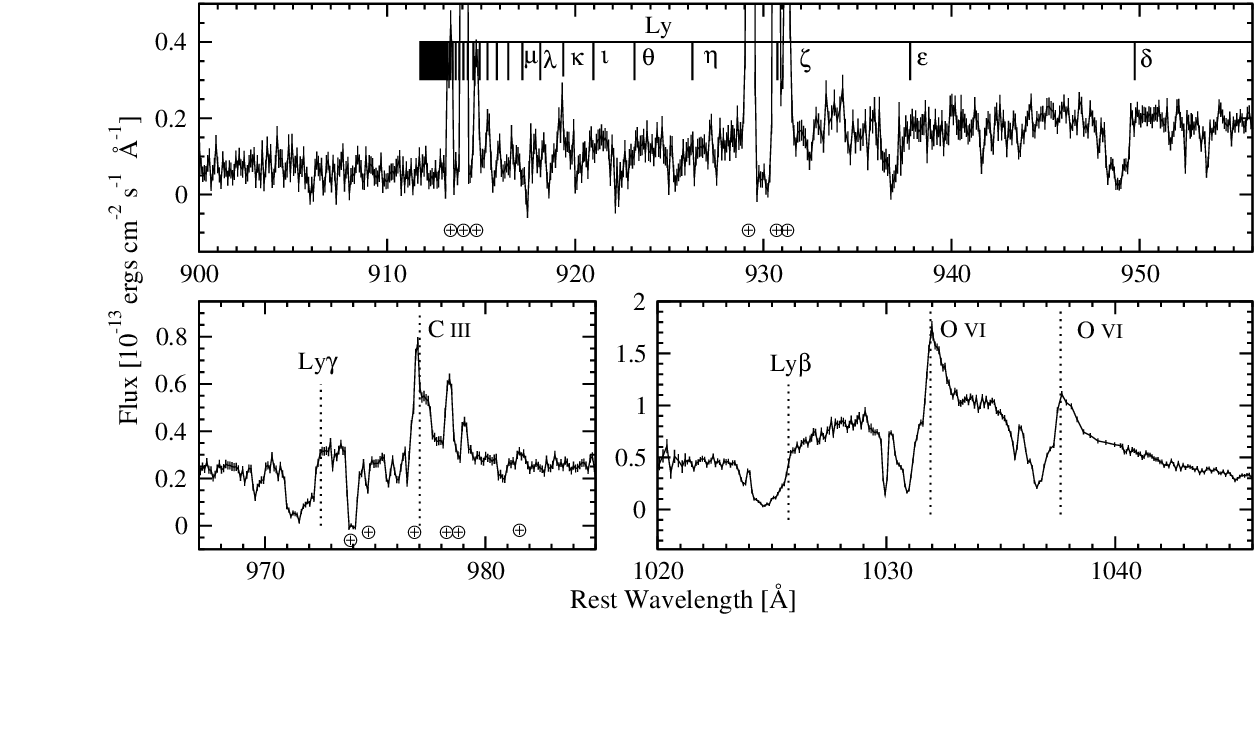}}
\vskip -0.65in
\caption{Selected bands from the {\it FUSE} spectrum of \mr . 
Identified intrinsic emission lines are marked at their
theoretical position with vertical lines and identified airglow
and Galactic lines are marked below the spectrum with~$\oplus$.}
\vskip -0.15in
\end{figure}

\section{The High Resolution {\it FUSE} spectrum}

The {\it FUSE} spectrum (Figure 3) shows broad emission lines of
O\,{\sc vi} and C\,{\sc iii} which also show significant blueshifted
absorption.  We also detect blueshifted absorption from at least
10 lines of the H\,{\sc i} Lyman series. We identify at least 4
absorption systems, one at $-$580 \kms\ and at least 3 others which
are blended together and form a wide trough covering the velocity
range of 0 to $-$500 \kms . The trough profiles are consistent with
the absorption in
the high resolution X-ray spectrum (though the later does not have
sufficient resolution) and suggest that the UV and X-ray absorptions
may arise from the same region in the AGN.

\acknowledgments

We acknowledge a financial support by the Israel Science Foundation
grant no. 545/00. S. K. also acknowledge financial support by Colton Foundation.


\begin{references}

\reference Ganguly, R., Charlton, J.~C., \& Eracleous, M.\ 2001, \apjl, 556, L7 

\reference Halpern, J.~P.\ 1984, \apj, 281, 90

\reference Mineo, T.~\& Stewart, G.~C.\ 1993, \mnras, 262, 817 

\reference Morales, R.~\& Fabian, A.~C.\ 2002, \mnras, 329, 209 

\reference Orr, A. et al. 2001, \aap, 376, 413 

\end{references}
\end{document}